# Long-range order induced by random fields in two-dimensional *O(n)* models, and the Imry-Ma state

A.A. Berzin[a], A.I. Morosov[b]*, and A.S. Sigov[a]

[a]MIREA - Russian Technological University, 78 Vernadskiy Ave., 119454 Moscow, Russian Federation

[b]Moscow Institute of Physics and Technology (National Research University), 9 Institutskiy per., 141700 Dolgoprudny, Moscow Region, Russian Federation

* e-mail: mor-alexandr@yandex.ru

## Abstract

The influence of defects of the "random local field" type with an anisotropic distribution of random fields on two-dimensional models with continuous symmetry of the vector order parameter is considered. In the case of weak anisotropy of random fields, with decreasing temperature there takes place a smooth transition from the paramagnetic phase with dynamic fluctuations of the order parameter to the Imry-Ma phase with static fluctuations caused by fluctuations of the random field of defects. In the case of strong anisotropy of random fields, defects lead to an effective decrease in the number of components of the order parameter and the appearance of a phase transition to an ordered state at finite temperature. It is shown that in the case of the defect-free two-dimensional *X-Y* model, the appearance of an arbitrarily weak anisotropy in the two-dimensional space of the order parameter completely eliminates the appearance of the Berezinsky-Kosterlitz-Thouless phase and gives rise to the phase with the long-range order.

## 1. Introduction

It is well known that in two-dimensional systems with continuous symmetry of the *n*-component vector order parameter ($O(n)$ model), a long-range order at finite temperature is absent. In the two-dimensional *X-Y* model ($n = 2$), at finite temperature, a phase transition occurs from the paramagnetic phase with the exponentially decreasing correlation function of the order parameter to the Berezinsky-Kosterlitz-Thoules (BKT) phase with the power-law character of the decrease of the correlation function [1-3]. In the two-dimensional Heisenberg model ($n = 3$), the paramagnetic phase exists at all nonzero temperatures [3]. The introduction of a weak single-ion anisotropy at low temperature leads to a change in the critical behavior of the system. In the case of anisotropy of the "easy axis" type, a phase transition to the ferromagnetic phase takes place [4]. In the case of the "easy plane" type anisotropy, a phase transition occurs from the paramagnetic phase to the BKT phase. A similar consideration of the influence of a weak anisotropy of the "easy axis" type on the behavior of the two-dimensional *X-Y* model was not carried out.

As shown in our papers [5–7], introduction into the system with $O(n)$ symmetry of the order parameter the "random local field" type defects with an anisotropic distribution of the directions of the random fields of defects in the *n*-dimensional space of the order parameter induces the effective global anisotropy to the second order of the random field. Such an anisotropy seeks to orient the order parameter perpendicular to the preferred direction of random fields. It is this anisotropy that explains the appearance of long-range order at finite temperature in the two-dimensional *X-Y* model as a result of the introduction of defects with collinear directions of local fields into the system [8, 9].

In addition, defects of the "random local field" type can significantly change the structure of the paramagnetic phase, transforming it into the inhomogeneous Imry-Ma state [10], in which the order parameter follows spatial fluctuations in the direction of the random field created by the defects.

This paper is devoted to constructing a phase diagram of the two-dimensional *X-Y* and Heisenberg models with the weak anisotropy induced by an anisotropic distribution of random fields of the defects.

## 2. System of classical spins

The exchange interaction energy of *n*-component localized spins $\mathbf{s}_i$ of a fixed unit length (the length of the vector can be included in the corresponding interaction constants or fields) forming a two-dimensional square lattice in the approximation of the interaction of nearest neighbors has the form

$$W_{ex} = -\frac{1}{2} J \sum_{i,\delta} \mathbf{s}_i \mathbf{s}_{i+\delta}, \qquad (1)$$

where *J* is the exchange integral, the summation over *i* is carried out over the entire lattice of spins, and over *δ* is carried out over the nearest neighbors to given spin.

In the presence of single-ion anisotropy in the system, the corresponding energy can be written in the form

$$W_{an} = \frac{1}{2} K \sum_i (s_i^\alpha)^2, \qquad (2)$$

where *K* is the anisotropy constant, $s_i^\alpha$ is one of *n* components of the *i*-th spin, and the summation over *i* is performed over the entire lattice of spins.

The interaction energy of spins with random local fields of defects is

$$W_{def} = -\sum_l \mathbf{s}_l \mathbf{h}_l, \qquad (3)$$

the summation is performed over randomly distributed defects, $\mathbf{h}_l$ is the local field of the *l*-th defect, and the distribution density of random local fields $\mathbf{h}$ in the spin space (the space of the order parameter) has the property $\rho(\mathbf{h}) = \rho(-\mathbf{h})$, which ensures the absence of a mean field in the infinite system.

## 3. Defect-free two-dimensional Heisenberg model with weak anisotropy

This problem was considered in Ref. [4]. In the case of the "easy axis" type ($K<0$) anisotropy, the system goes over to the class of Ising models and the long-range order arises in it at temperature $T_c$ equal to [4]

$$T_c \approx \frac{4\pi J}{\ln\left(\frac{J}{|K|}\right)}. \tag{4}$$

At $T \sim T_c$, the number $n$ decreases effectively from three to unity, and the long-range order arises in the system.

Expression (4) is easy to obtain from simple energy considerations. To do this, we equate by the absolute value the temperature $T_c$ and the anisotropy energy $E_c \sim K\xi^2$ of the region correlated at $T < J$ with a radius equal to the correlation radius $\xi$. The correlation radius $\xi$ in the two-dimensional Heisenberg model is [3]

$$\xi = \exp\left(\frac{2\pi J}{T}\right). \tag{5}$$

Solving the resulting equation for the variable $T_c$

$$T_c = |K| \exp\left(\frac{4\pi J}{T_c}\right) \tag{6}$$

by iteration procedure and neglecting a numerical factor of the order of unity under the logarithm, we obtain, in a first approximation, formula (4). This method of estimating $T_c$ also works well when studying systems with a strongly anisotropic exchange interaction, for example, to find the critical temperature of the quasi-one-dimensional Ising model [11]. In this case, it is necessary to equate to $T_c$ not the anisotropy energy, but the energy of the weak exchange interaction between the correlated one-dimensional regions of neighboring filaments, along which a strong exchange interaction takes place. If a weak anisotropy of the "easy plane" type arises in the two-dimensional Heisenberg model, then at temperature $T_c$ given by formula (4), a phase transition to the BKT phase occurs [4], and the effective number of components of the order parameter decreases from three to two.

## 4. Defect-free two-dimensional *X-Y* model with weak anisotropy

In this model with *n*=2, only an anisotropy of the "easy axis" type is possible. We will also use the $T_c$ estimation method described in the previous section for this model. According to Ref. [3] one has

$$\xi = \exp(b\tau^{-1/2}), \qquad (7)$$

where $\tau = (T - T_{BKT})/T_{BKT}$, and $T_{BKT} = \pi J/2$ is the transition temperature from the paramagnetic phase to the BKT phase in the absence of anisotropy. From the condition $|E_c| \sim T_c$, we find, to a first approximation, assuming that $T_c \sim T_{BKT}$,

$$\tau_c \approx \frac{4b^2}{\ln^2 \frac{J}{|K|}} \ll 1. \qquad (8)$$

Thus, the presence of a weak anisotropy leads to the fact that at temperature $T_c = (1 + \tau_c)T_{BKT}$, a phase transition to the phase with the long-range order occurs in the system, and the BKT phase does not occur. Since a weak anisotropy induced by the symmetry of the crystal lattice is always present in real crystalline systems, experimental observation of the BKT phase in such systems is hardly possible.

## 5. Temperature $\tilde{T}$ of the Imry-Ma phase occurrence

In contrast to the systems considered in Sections 3 and 4 where the single-ion anisotropy was present in a defect-free system, this section will consider the case when anisotropy is induced by the defects of the "random local field" type, that is, only the terms (1) and (3) are present in the energy of classical spins. But these defects generate not only the anisotropy, but also random field fluctuations, which contribute to the destruction of the long-range order. The transition from the behavior inherent in the paramagnetic phase of the pure system to the behavior characteristic of the disordered Imry-Ma phase occurs at temperature $\tilde{T}$, which is found from the condition that the correlation radius of the pure system $\xi$ is equal to the optimal length $L^*$ of static fluctuations of the order parameter in the Imry-Ma phase. For $T > \tilde{T}$, the condition $\xi < L^*$ is fulfilled, and dynamic thermal fluctuations of the order parameter characteristic of a pure system are observed in the system. For $T < \tilde{T}$, the relation $\xi > L^*$ is valid and the static fluctuations of the random field "freeze" these dynamic fluctuations. Such a phenomenon is akin to the glass transition phenomenon.

According to Ref [12], in the case of the two-dimensional system we have in the order of magnitude

$$L^* \approx \left(\frac{J^2}{x \langle \mathbf{h}_l^2 \rangle}\right)^{1/2}. \tag{9}$$

Here $x$ is the dimensionless concentration of the defects (their number per a unit cell), and the angle brackets denote averaging over the fields of all defects. In the case of the two-dimensional Heisenberg model, from the condition $\xi = L^*$ we obtain an estimate for the quantity $\tilde{T}$

$$\tilde{T} \approx \frac{2\pi J}{\ln L^*} \approx \frac{4\pi J}{\ln\left(\frac{J^2}{x \langle \mathbf{h}_l^2 \rangle}\right)} = \frac{4\pi J}{\ln\left(\frac{J}{K_{cr}}\right)}, \tag{10}$$

where $K_{cr} \sim x \langle \mathbf{h}_l^2 \rangle / J$ is the critical value of the global anisotropy of the "easy axis" type, which makes the existence of the Imry-Ma phase energetically disadvantageous [12].

In the case of the two-dimensional *X-Y* model, from the same condition we find the value $\tau$ corresponding to the fluctuations freezing

$$\tilde{\tau} \approx \frac{4b^2}{\ln^2\left(\frac{J^2}{x \langle \mathbf{h}_l^2 \rangle}\right)} \approx \frac{4b^2}{\ln^2\left(\frac{J}{K_{cr}}\right)}. \tag{11}$$

The temperature of the transition to the Imry-Ma state $\tilde{T} = (1 + \tilde{\tau})T_{BKT} \sim T_{BKT}$ itself is close to the transition temperature in the pure system. A similar situation occurs in three-dimensional *O(n)* models, where $\tilde{T}$ is of the order of the temperature of the appearance of the long-range order in the pure system.

## 6. Phase diagram of the two-dimensional Heisenberg model with defects of the "random local field" type

Comparing formulas (4) and (10), it is easy to see that the condition for suppressing the Imry-Ma disordered state $|K| > K_{cr}$ obtained in [12] is equivalent to the condition $T_c > \tilde{T}$. Otherwise, with decreasing temperature at $T \sim \tilde{T}$, a transition from the paramagnetic phase to the Imry-Ma phase occurs, which is described in the previous section. Consider the phase diagram in the region $T_c > \tilde{T}$.

With decreasing temperature, at $T = T_c$, in the case of the defect-induced anisotropy of the "easy axis" type, a transition to the adequately one-component order parameter takes place, and in the case of the defect-induced anisotropy of the "easy plane" type, a transition to the adequately two-component order parameter occurs. Since the lowest critical dimensionality for the Ising model with defects of the "random local field" type equals two [13] (see also the review paper [14]), a transition to the one-component order parameter in the two-dimensional system is accompanied by the ferromagnetic phase initiation, that is, the paramagnetic-ferromagnetic phase transition is set of.

If at $T = T_c$ the transition proceeds to the adequately two-component order parameter, then for an investigation of the behavior of the system originated one needs to project the random fields of defects onto the easy plane, to calculate the parameters $|K'|$, $K'_{cr}$, $T'_c$, and $\tilde{T}'$ for the system with a less number of the order parameter components, and to study the part of the phase diagram of the *X-Y* model with the "random local field" type defects related to $T < T_c$ temperature range.

In Ref. [7] there was obtained the self-consistency equation for the global effective anisotropy created by the anisotropic distribution of random fields in the space of the two-dimensional system order parameter.

$$|K| = \frac{x}{4\pi J}\left(p^{max} - p^{min}\right)\ln\frac{4\pi J}{|K|}, \qquad (12)$$

where $p$ is the quadratic form with respect to the components of the unit vector $\mathbf{s}_0$ parallel to the mean spin direction,

$$p = \sum_{\alpha=1}^{n} s_{0\alpha}^2 \langle h_{l\alpha}^2 \rangle, \tag{13}$$

and $p^{max}$, $p^{min}$ are maximum and minimum values of this form with respect to all possible orientations $\mathbf{s}_0$ in the n-dimensional space of the order parameter.

In the case of the coplanar and isotropic in the selected plane distribution of random fields, the easy axis arises perpendicular to this plane. The self-consistency equation for the anisotropy constant (12) takes the form [7]

$$|K| = \frac{x \langle \mathbf{h}_l^2 \rangle}{8\pi J} \ln \frac{4\pi J}{|K|}. \tag{14}$$

Solving it by the iteration method, we obtain in the first approximation

$$|K| = \frac{x \langle \mathbf{h}_l^2 \rangle}{8\pi J} \ln \frac{32\pi^2 J^2}{x \langle \mathbf{h}_l^2 \rangle}. \tag{15}$$

In the case $T_c > \tilde{T}$, the order parameter in the ferromagnetic phase is oriented collinearly to the easy axis, its large-scale static fluctuations are suppressed, only local deviations of the spins near the defects occur. For $T_c < \tilde{T}$, the spins in the Imry-Ma phase lie in the same plane as random fields.

The condition $T_c > \tilde{T}$ for $J^2/\langle \mathbf{h}_l^2 \rangle = 100$ corresponds to the concentration range $x < x_c = 4 \cdot 10^{-7}$. Thus, even with a given distribution of the random fields directions, optimal for creating the global anisotropy, the Imry-Ma phase will be observed for all actual concentrations of defects. The phase diagram of the two-dimensional Heisenberg model with defects of the "random local field" type and anisotropy of the "easy axis" type is shown in Fig. 1.

In the case when all random fields are collinear to the $z$ axis of the Cartesian orthogonal coordinate system, the easy plane $xy$ appears in the spin space. As this takes place, the projections of random fields on the easy plane are equal to zero, which makes the resulting system with the two-component order parameter equivalent to the pure X-Y model. Therefore, at the temperature $T_c$ given by formula (4) with the value $/K/$ equal to [7]

$$|K| = \frac{x\langle \mathbf{h}_l^2 \rangle}{4\pi J} \ln \frac{16\pi^2 J^2}{x\langle \mathbf{h}_l^2 \rangle}, \tag{16}$$

there takes place the phase transition to the BKT phase, in which the spins lie in the easy plane.

The condition $T_c > \tilde{T}$ for $J^2/\langle \mathbf{h}_l^2 \rangle = 100$ corresponds to the concentration range $x < x_c = 5.5 \cdot 10^{-2}$. Thus, the transition from the BKT phase to the Imry-Ma phase with increasing defect concentration occurs at a noticeable critical concentration. The phase diagram of the two-dimensional Heisenberg model with defects of the "random local field" type and anisotropy of the "easy plane" type is shown in Fig. 2.

## 7. Phase diagram of the two-dimensional *X-Y* model with defects of the "random local field" type

Similarly to the consideration performed in the previous section, the condition for suppressing the Imry-Ma disordered state $|K| > K_{cr}$ is equivalent to the relation $\tau_c > \tilde{\tau}$. If $|K| < K_{cr}$, then for $\tau \sim \tilde{\tau}$ the paramagnetic phase goes over to the Imry-Ma state.

In the case of an anisotropic distribution of the directions of random fields, for which all $\mathbf{h}_l$ are collinear, the easy axis perpendicular to the direction of random fields appears in the system, and the value of the anisotropy constant is given by formula (15). Since the projections of random fields on the easy axis are equal to zero, the resulting system with the one-component order parameter is equivalent to the pure two-dimensional Ising model. Therefore, at $\tau = \tau_c$, the system undergoes the phase transition from the paramagnetic phase to the ferromagnetic one, whose structure is similar to the structure of the ferromagnetic phase described in the previous section. The condition $\tau_c > \tilde{\tau}$ is fulfilled in the concentration range $x < x_c = 5.5 \cdot 10^{-2}$. However, even at high concentrations, when $\tau_c < \tilde{\tau}$ and the disordering effect of random fields prevails, the long-range order exists in the system. The reasons for this were considered in our paper [15].

Since all random fields are collinear with some straight line in the two-dimensional space of the order parameter (we choose it as the $\xi$ axis of the Cartesian orthogonal coordinate system ($\xi, \eta$)), the fluctuation of the random field in a region with a characteristic linear size $L^*$ is either parallel or antiparallel to the axis $\xi$. When passing from a region with one direction of the field to a neighboring region with the opposite direction of the field, the order parameter rotates through 180°. The energy of the inhomogeneous exchange will be obviously lower if this turn in the entire spin lattice occurs along the same arc passing, for example, through the positive semiaxis $\eta$. In this case, the nonzero average value of the vector order parameter parallel to the $\eta$ axis appears in the system. This explains the results of modeling [8, 9], where the random field was introduced in each unit cell ($x=1$) and the presence of the long-range order induced by random fields (random field induced order) was demonstrated. The average value of the order parameter, even in the ground state, is far from saturation, since fluctuations of the random field lead to significant static fluctuations of the order parameter. The coexistence of the Imry-Ma phase and the ferromagnetic phase takes place. It should be noted that such a complex phase seems to be unique. In the entire range of defect concentrations, the phase transition to the ferromagnetic phase or the Imry-Ma phase with the long-range order occurs at $T \sim T_{BKT}$. The phase diagram of the two-dimensional *X-Y* model with defects of the "random local field" type is shown in Fig. 3.

## 8. Conclusions

1. Anisotropic distribution of random fields of defects, creating a global anisotropy, reduces adequately the number of components of the order parameter. In the case of two-dimensional *O(n)* models, this can lead to the appearance of the long-range order in the system at finite temperature.
2. In the case of the defect-free two-dimensional *X-Y* model, the appearance of an arbitrarily weak anisotropy in the two-dimensional space of the order parameter completely eliminates the appearance of the Berezinsky-Kosterlitz-Thouless phase and gives rise to the phase with the long-range order.

3. Since defects of the "random local field" type create static fluctuations of the field that contribute to disordering of the spin system, in the two-dimensional Heisenberg model with the given defects and the "easy axis" type anisotropy at the defect concentration exceeding $4 \cdot 10^{-7}$, with lowering temperature, a smooth transition occurs from dynamic fluctuations of the order parameter characteristic of a pure system to the Imry-Ma phase with static fluctuations of the order parameter following fluctuations of the random field direction.

4. In the case of the two-dimensional *X-Y* model and collinear random defect fields, at temperature $T \sim T_{BKT}$ and the defect concentration lower than $5.5 \cdot 10^{-2}$, the phase transition from the paramagnetic phase to the ferromagnetic phase occurs, and at a higher defect concentration, the paramagnetic phase passes to the Imry-Ma phase with the long-range order.

# References

1. V. L. Berezinskii, Sov. Phys. JETP **32**, 493 (1971); **34**, 610 (1972).
2. J. M. Kosterlitz, D. G. Thouless. J. Phys. C **6**, 1181 (1973).
3. Yu.A. Izyumov, Yu.N. Skryabin. Statistical mechanics of magnetically ordered systems. Nauka, Moscow (1987). 268 c. (in Russian).
4. S.B. Khokhlachev, Sov. Phys. JETP **43**, 137 (1976).
5. A.A. Berzin, A.I. Morosov, A.S. Sigov, Phys. Solid State **58**, 1671 (2016).
6. A.A. Berzin, A.I. Morosov, A.S. Sigov, Phys. Solid State **58**, 1846 (2016).
7. A.A. Berzin, A.I. Morosov, A.S. Sigov, Phys. Solid State **59**, 2016 (2017).
8. B. J. Minchau, R. A. Pelcovits. Phys. Rev. B **32**, 3081 (1985).
9. J. Wehr, A. Niederberger, L. Sanchez-Palencia, M. Lewenstein. Phys. Rev. B **74**, 224448 (2006).
10. Y. Imry, S.-k. Ma. Phys. Rev. Lett. **35**, 1399 (1975).
11. D. J. Scalapino, Y. Ymry, P. Pincus. Phys. Rev. B **11**, 2042 (1978).
12. A.I. Morosov, A.S. Sigov, JETP Letters **90**, 723 (2009).
13. J. Imbrie. Phys. Rev. Lett. **53**, 1747 (1984).
14. V.S. Dotsenko, Physics-Uspekhi **38,** 457 (1995).
15. A.A. Berzin, A.I. Morosov, Phys. Solid State **57**, 2217 (2015).

## Figure Captions

1. Phase diagram of the two-dimensional Heisenberg model with the "easy axis" type anisotropy induced by the "random local field" type defects, for $J^2/\langle \mathbf{h}_l^2 \rangle = 100$: $P$ is the paramagnetic phase, $F$ is the ferromagnetic phase, and $I$-$M$ is the disordered Imry-Ma phase.

2. Phase diagram of the two-dimensional Heisenberg model with the "easy plane" type anisotropy induced by the "random local field" type defects, for $J^2/\langle \mathbf{h}_l^2 \rangle = 100$: $P$ is the paramagnetic phase, $BKT$ is the BKT phase, and $I$-$M$ is the disordered Imry-Ma phase.

3. Phase diagram of the two-dimensional $X$-$Y$ model with the collinear directions of the random fields of the defects, for $J^2/\langle \mathbf{h}_l^2 \rangle = 100$: $P$ is the paramagnetic phase, $F$ is the ferromagnetic phase, and $I$-$M$ + $LRO$ is the Imry-Ma phase with the long-range order.

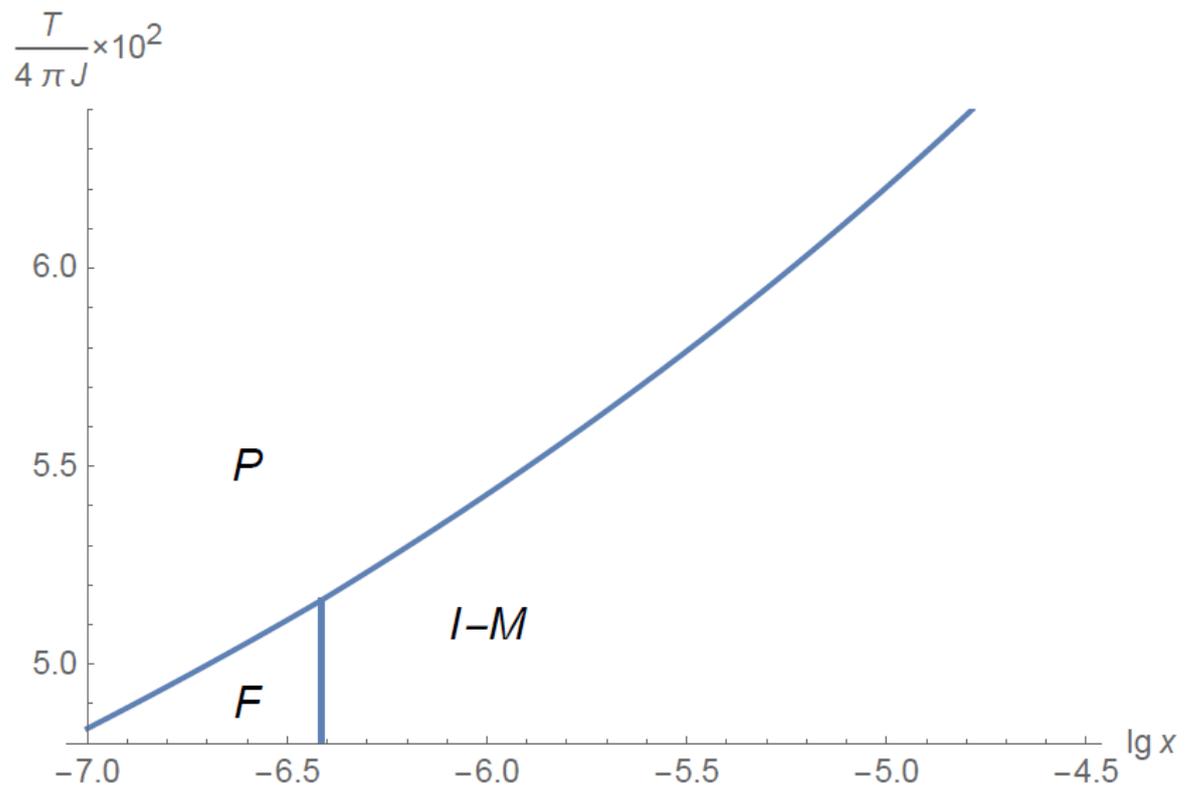

Fig. 1

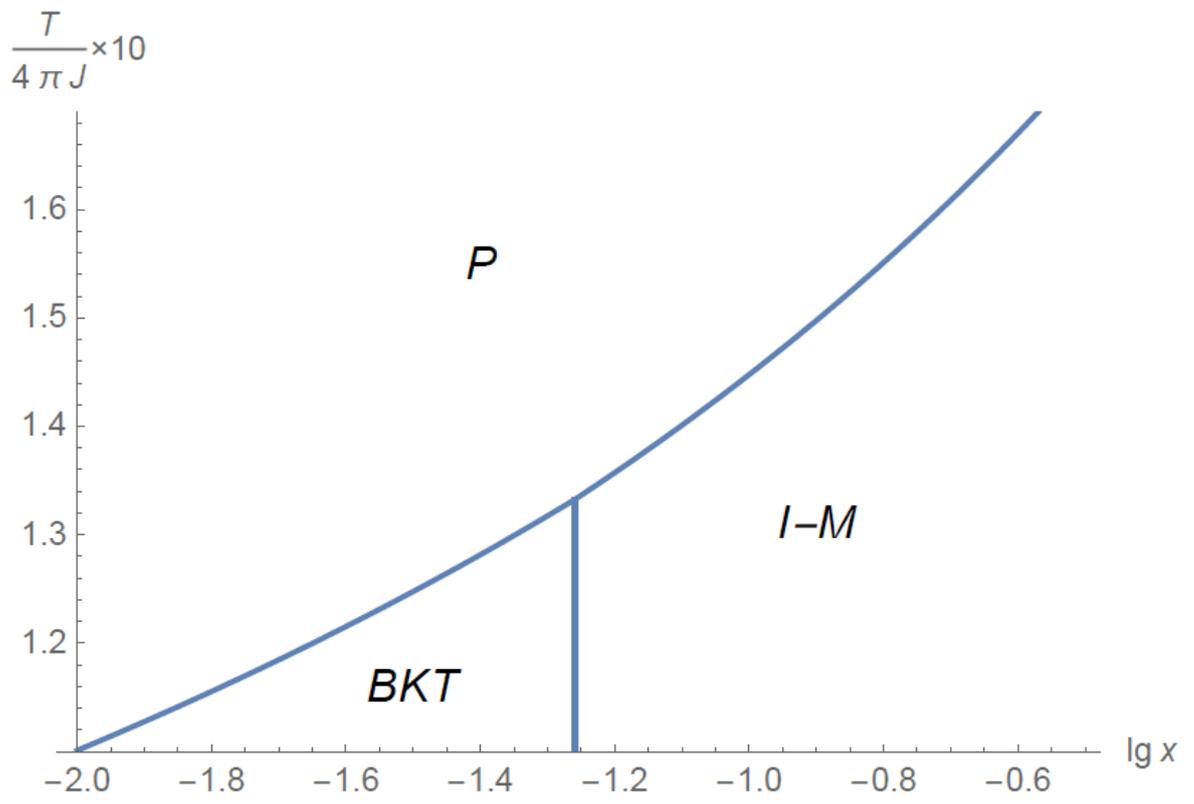

Fig. 2.

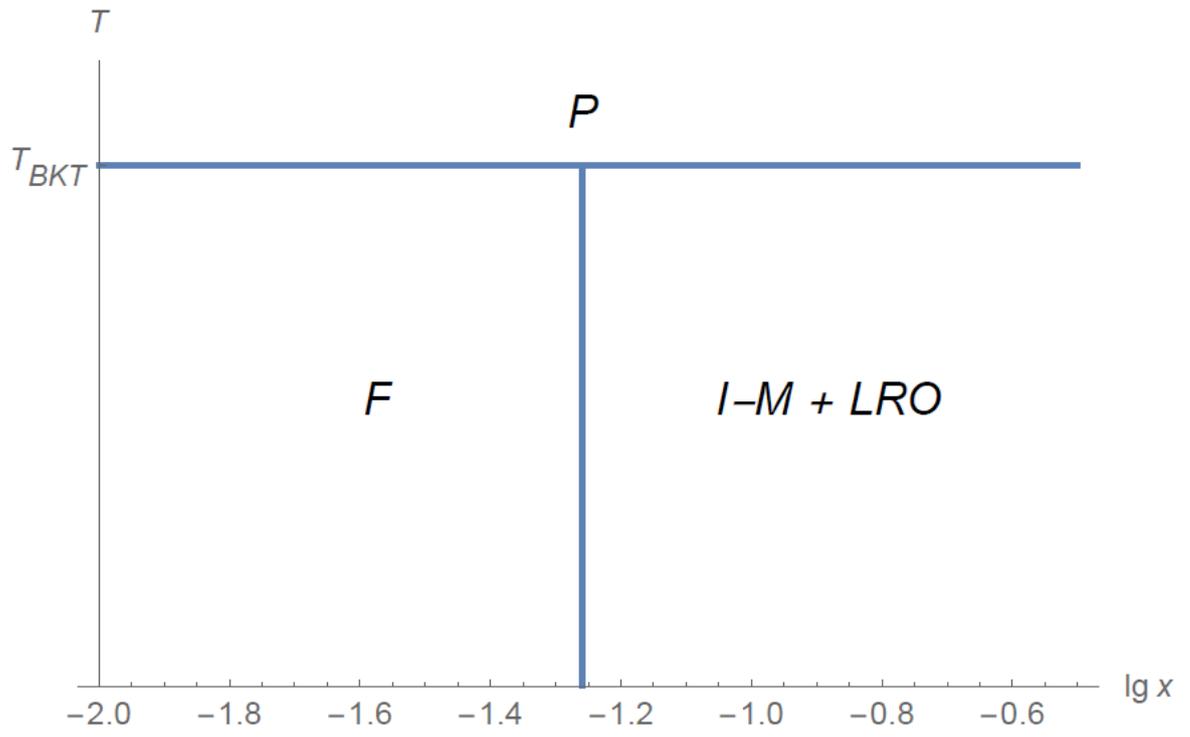

Fig. 3.